\documentclass[notitlepage,twocolumn,prl]{revtex4-1}

\usepackage{amsmath}
\usepackage{changes}
\usepackage{graphicx}
\usepackage{changes}
\usepackage{soul}
\usepackage{blindtext}
\usepackage[colorlinks=true,citecolor=blue]{hyperref}

\begin{document}

\title{Dynamics of soliton crystals in optical microresonators}

\author{Maxim Karpov}
\author{Martin H. P. Pfeiffer}
\author{Hairun Guo}
\author{Wenle Weng}
\author{Junqiu Liu}
\author{Tobias J. Kippenberg}
\email{tobias.kippenberg@epfl.ch}
\affiliation{Swiss Federal Institute of Technology in Lausanne (EPFL), CH-1015, Switzerland}

\date{\today}

\begin{abstract}
Dissipative Kerr solitons in optical microresonators provide a unifying framework for nonlinear optical physics with photonic-integrated technologies and have recently been employed in a wide range of applications from coherent communications to astrophysical spectrometer calibration. Dissipative Kerr solitons can form a rich variety of stable states, ranging from breathers to multiple-soliton formations, among which, the recently discovered soliton crystals stand out. They represent temporally-ordered ensembles of soliton pulses, which can be regularly arranged by a modulation of the continuous-wave intracavity driving field. To date, however, the dynamics of soliton crystals remains mainly unexplored. Moreover, the vast majority of the reported crystals contained defects - missing or shifted pulses, breaking the symmetry of these states, and no procedure to avoid such defects was suggested. Here we explore the dynamical properties of soliton crystals and discover that often-neglected chaotic operating regimes of the driven optical microresonator are the key to their understanding. In contrast to prior work, we prove the viability of deterministic generation of \emph{perfect} soliton crystal states, which correspond to a stable, defect-free lattice of optical pulses inside the cavity. We discover the existence of a critical pump power, below which the stochastic process of soliton excitation suddenly becomes deterministic enabling faultless, device-independent access to perfect soliton crystals. Furthermore, we demonstrate the switching of soliton crystal states and prove that it is also tightly linked to the pump power and is only possible in the regime of transient chaos. Finally, we report a number of other dynamical phenomena experimentally observed in soliton crystals including the formation of breathers, transitions between soliton crystals, their melting, and recrystallization.
\end{abstract}
\pacs{}

\maketitle

\label{Introduction}
\noindent \textbf{Introduction.} --- Dissipative Kerr solitons (DKS) are optical pulses generated in Kerr-nonlinear optical resonators, which rely on a double balance between the dispersion and the nonlinearity of the cavity medium as well as the continuous-wave (CW) parametric gain and cavity losses \cite{Haelterman1992MI,Akhmediev2003,kippenberg2018DKS,herr2014soliton}. DKS generated in optical microresonators have recently attracted significant attention as a source of broadband, low-noise optical frequency combs with a number of unique properties such as chip-scale footprint, high repetition rates covering microwave to terahertz domains and CMOS-compatibility, enabling their wafer-scale production and photonic integration \cite{herr2014soliton,Brasch2014Cherenkov,raja2018electricalMicrocomb,stern2018batterymicrocomb}. The promising potential of microresonator-based DKS has been already demonstrated in a wide range of cutting-edge applications, spanning from optical coherent communications to the calibration of astronomical spectrometers \cite{marin2016DKScommunication, Jost2014link, dualcomb2016vahala, yu2016DKSdualcombMIR, trocha2017dualcombLidar, suh2018lidarVahala, liang2015DKSlowRF, spencer2018DKSsyntetizer, obrzud2017microphotonic,suh2018exoplanetMicrocombVahala}. In addition, such CW-driven nonlinear microresonators are of high fundamental interest, as they provide a testbench for the exploration of spatiotemporal light localization and dynamics of nonlinear systems.

Apart from bright dissipative Kerr solitons, whose investigation has been a main focus since their experimental realization in microresonators, a rich variety of other stable waveforms including platicons (dark pulses) \cite{xue2015darksol,Lobanov2015platicons}, and bright and dark breathers \cite{Matsko2012breathers,bao2016FPU,lucas2017breathing,yu2017breathers,bao2017darkDKSbreathers,Leo2013} have been observed. Moreover, the behaviour of all these structures can be impacted by various effects, such as Raman scattering \cite{milian2015interplayRamanTOD,karpov2015raman,yang2017stokesDKS}, higher-order microresonator dispersion and avoided modal crossings \cite{Herr2014avoidmodecross,Brasch2014Cherenkov,guo2017inter-modeBreathers,yi2017SMdispersiveWave,lucas2017detuning,karpov2018photonic}, and thermal nonlinearity \cite{karpov2016universal}, giving rise to unanticipated nonlinear dynamics.

\begin{figure*}
\includegraphics[width = \textwidth]{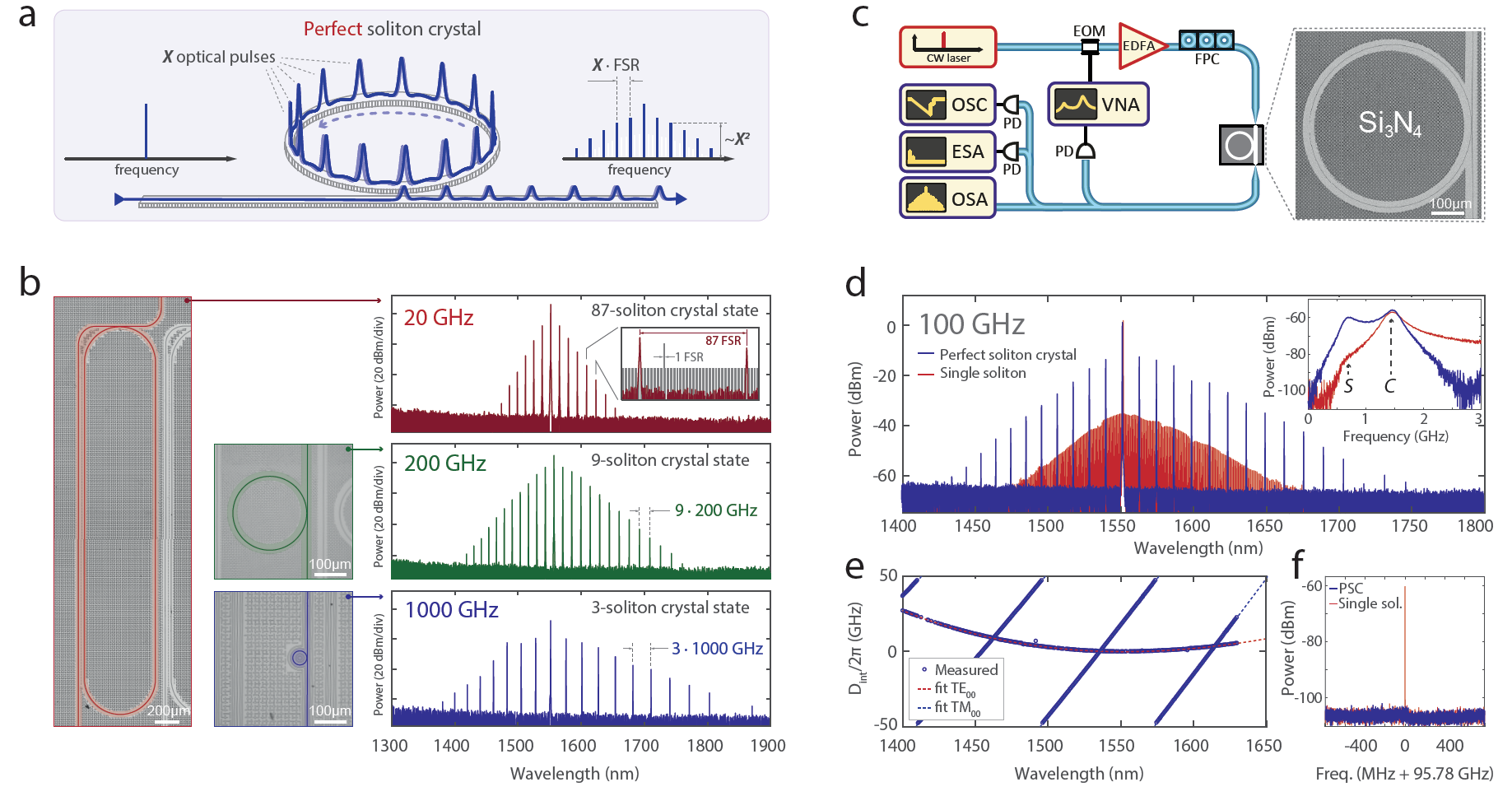}
\protect\caption{\textbf{Perfect soliton crystals in \boldmath{$\rm Si_3N_4$} microresonators}
(a) Sketch of the perfect soliton crystal consisting of $X$ pulses formed in the CW-driven  nonlinear optical microcavity.
(b) Optical microscope images of $\rm Si_3N_4$ microresonators with different free spectral ranges of 20 GHz, 200 GHz and 1000 GHz (left), and perfect soliton crystal states generated in each device (right); 
(c) Left: Set-up scheme used for the generation and characterization of DKS crystal states: tunable external-cavity diode laser with a center wavelength of 1550 nm is used as a seed, EDFA -- erbium-doped fiber amplifier, FPC -- fiber polarization controller, VNA -- vector network analyzer, EOM -- electro-optical phase  modulator, PD -- photodioide, OSC -- oscilloscope, ESA -- electrical spectrum analyzer, OSA -- optical spectrum analyzer; Right: optical microscope image of a 100-GHz $\rm Si_3N_4$ microresonator; 
(d) Optical spectra of the soliton crystal state (blue) and single soliton state (red) stabilized in a $\rm Si_3N_4$ microresonator device shown in (c) under the same conditions of pump power and effective detuning; inset shows the system response measurement using a VNA-based scheme \cite{karpov2016universal} in both states. \emph{C} and \emph{S} letters indicate the positions of $\mathcal{C}$- and $\mathcal{S}$-resonances correspondingly.
(e) Measured integrated dispersion of the $\rm Si_3N_4$ microresonator shown in (c) (circles) and fitting curves for the fundamental TE (dashed red) and TM (dashed blue) mode families \cite{Delhaye2009disp}. The calculated group velocity dispersion term ($D_{2}/2\pi$) for the TE mode used for DKS formation is $\sim 1.2$ MHz;
(f) Repetition rate beatnote measurements in the single soliton state and the PSC state shown in (d). Native repetition rate of $\sim 100$ GHz is undetectable in the PSC state;
 \label{fig_1}}
\end{figure*}  
Recent work has demonstrated that bright dissipative Kerr solitons are able to form temporally-ordered ensembles -- soliton crystals \cite{cole2016solitoncrystal}. Their optical spectra are characterized by a set of strongly enhanced comb lines spaced apart by multiple free spectral ranges (FSR) resulting from the regular arrangement of DKS pulses in the cavity.  The formation of such structures was shown to be linked to the presence of avoided modal crossings (AMX) \cite{cole2016solitoncrystal}, which through the spectrally-localized alterations of the microresonator dispersion changes the optical spectrum of a DKS state and induces a modulation on the CW intracavity background, leading to the ordering of DKS pulses in a crystal-like structure \cite{wang2017universalbindingDKS,taheri2017opticalLatticeTrap,cole2016solitoncrystal}.

Yet to date, the dynamics of such soliton crystal states remains mainly unexplored. While several attempts have been made to explore their defect morphology \cite{cole2016solitoncrystal,wang2018robust}, stability chart \cite{Karpov2017solitonCrystal, wang2018robust} and impact of Raman effects \cite{lu2018raman}, there is a number of open fundamental questions regarding the conditions of their excitation, impact of the chaotic operation regimes, operation stability and accessibility of defect-free states.  It is also unknown whether soliton crystals reproduce the typical behaviour of soliton states - if they feature similar switching mechanisms, or have enough robustness to form nonstationary states such as soliton crystal breathers.

Here we demonstrate the generation of \emph{perfect} soliton crystal states, which, strikingly, could be accessed in a deterministic fashion. More generally, for the first time we study the full range of dynamical properties of soliton crystals. First, we shed light on the formation process of these states, and demonstrate their critical dependency on the excitation pump power, which originates from two often-neglected operating regimes of the driven microresonator -- spatiotemporal chaos and transient chaos. We discover the existence of a critical pump power, below which the originally stochastic process of soliton excitation becomes deterministic and provides faultless access to perfect soliton crystals. Despite their similarity to primary combs, we unambiguously prove the presence of an underlying soliton crystal lattice, which gives rise to the perfect interference and selective enhancement and suppression of comb teeth in these states. Second, we demonstrate that soliton crystals can be reliably translated in the two-dimensional parameter space of the system (pump power and detuning). Using such translations we experimentally investigate their behaviour in different stability regimes and demonstrate that soliton crystal states are able to be switched. Importantly, we establish a fundamental link between the soliton switching phenomenon and the regime of transient chaos, which we prove to be solely responsible for DKS elimination. Finally, we present here a rich panel of experimentally-observed, novel dynamical phenomena appearing in soliton crystals, including the first observation of soliton crystal breathers, switching between perfect soliton crystals, as well as controllable soliton crystal melting, disordering and recrystallization.

\noindent \textbf{Perfect soliton crystals} --- 
\label{Perfect soliton crystals}
Despite a broad variety of soliton crystal states \cite{cole2016solitoncrystal}, here we mainly focus on their simplest and most ideal representatives - \emph{perfect} soliton crystals (PSC) \cite{Karpov2017solitonCrystal,wang2018robust}. 

A PSC is a set of dissipative Kerr solitons distributed evenly on the resonator circumference, whose number ($X$) equals to the maximum number of solitons that such resonator can accommodate under given pumping conditions. In contrast to soliton crystal states with defects, the behaviour of PSC is unperturbed by missing or shifted pulses, what guarantees an access to the "pure" dynamics of soliton crystals.
Besides their simplicity, the PSC state bears several important features relevant for both fundamental research and applications. First, due to the high regularity of the intracavity pulses, the PSC states can be used as high-purity, ultra-high-repetition rate soliton combs, reaching a mode spacing of several THz (which is challenging for small microresonators due to bending losses and limitations on the dispersion control). The second advantage of PSC states is that the comb energy is distributed in a few lines (supermodes), separated by $X\cdot$FSRs, which gives them an $X^2$ enhancement in comparison to the single-soliton state excited under the same conditions (see Fig.\ref{fig_1}(a,d)). The latter can be especially useful for self-referencing of such combs, as well as locking to an optical reference located in the weak wings of single-soliton states. 

In experiments we employ $\rm Si_3N_4$  microring resonators with various FSRs of 20, 100, 200 and 1000 GHz (see Fig.\ref{fig_1}(a, b)), fabricated with two different fabrication processes - subtractive \cite{luke2013overcomingSi3N4sress, Brasch2014Cherenkov} and photonic Damascene process \cite{Pfeiffer2015Damascene}. A typical setup for DKS excitation shown in Fig.\ref{fig_1}(c) includes an additional electro-optical modulator (EOM) and vector network analyzer (VNA) in order to measure the cavity response and probe the DKS state.

Fig.\ref{fig_1}(d) shows a particular PSC state with $X=15$, generated in a 100-GHz microresonator (blue) and the single soliton state (red) generated in the same device under exactly the same conditions (pump power and effective detuning). Despite an apparent similarity of the PSC states to primary combs, we unambiguously prove the soliton formation in a PSC state with response measurements (inset of Fig.\ref{fig_1}(d)), which show a clear double-resonance response indicating the coexistence of DKS pulse with CW background. Both generated soliton states (single soliton and PSC) operate at the same effective detuning (can be estimated by the position of the $\mathcal{C}$-resonance), while the number of pulses differs significantly (can be observed from the amplitude of the $\mathcal{S}$-resonance). A fascinating feature of the generated PSC states is the complete suppression of all the other comb lines apart from the supermodes(see Fig.\ref{fig_1}(f)), which corresponds to extreme regularity of the DKS pulse arrangement inside the cavity and confirms the absence of any defects. Furthermore, we could not detect the native-FSR beatnote of $\sim 100$ GHz electronically in the PSC states, and the observed difference between supermodes and the noise floor established by OSA was at least 60 dB.

\begin{figure*}
\includegraphics[width = \textwidth]{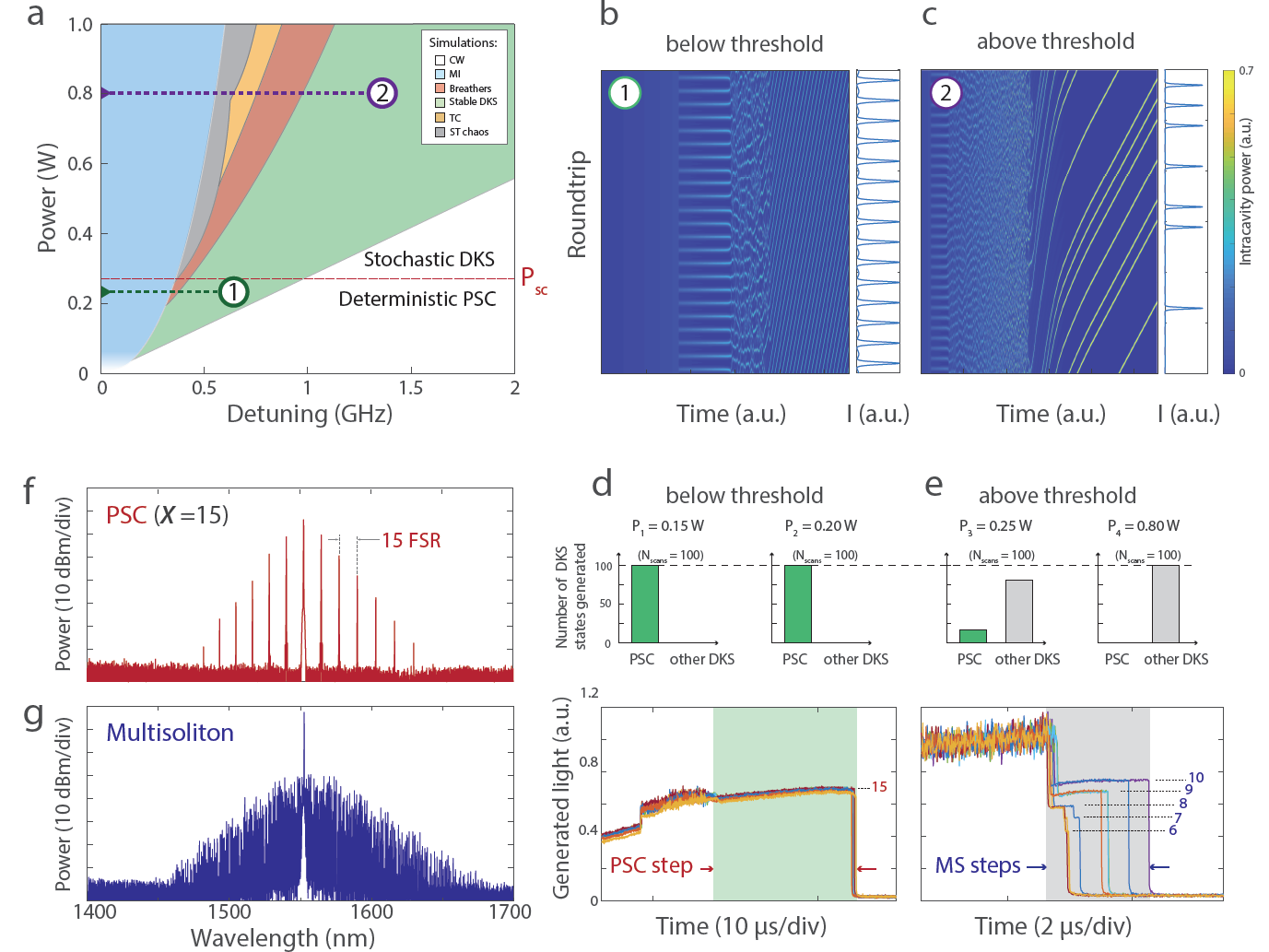}
\protect\caption{\textbf{Universal approach to the excitation of perfect soliton crystal states (PSC)}
(a) Simulated stability chart of the Lugiato-Lefever equation augmented with one AMX crossing (see Supplementary information for details); The coloured areas indicate different stability regions of the PSC states: stable PSC - green, breathing PSC - red, modulation instability - blue, spatiotemporal chaos (STC) - grey, transient chaos (TC) - yellow, CW solutions - white. Numbered dashed lines indicate two tuning procedures generating DKS states at fixed pump power which avoid (1), or go through the STC and TC regions (2). 
(b, c) Simulations of the intracavity waveform evolution during two tuning attempts for the generation of DKS states involving and avoiding TC and STC regions: 1 --- pump power $P_{\rm in} < P_{\rm sc}$, both STC and TC regions region are avoided, the detuning is stopped in the stable soliton regime - PSC state is generated; 2 --- $P_{\rm in} > P_{\rm sc}$ the detuning is stopped in the stable soliton regime - multiple-soliton state is generated. 
(d) Experimental pump sweeps over the cavity resonance at two pump powers $P_1 = 0.15$W and $P_2 = 0.20$W below the threshold power $P_{\rm sc}$; top: statistical overview of the generated states out of 100 scans for each power - at both powers the system \emph{always} demonstrate the step of the same height, corresponding to a PSC state with $X=15$; bottom: arbitrarily chosen five scan traces at $P_2$ showing PSC step reproducibility;
(e) Experimental pump sweeps over the cavity resonance at two pump powers $P_3 = 0.25$W and $P_4 = 0.80$W above the threshold power $P_{\rm sc}$; top: statistical overview of the generated states out of 100 scans for each power - no deterministic PSC step formation is observed, steps are formed stochastically, because the system passes through instability regions of STC or(and) TC; bottom: several scan traces at $P_4$ demonstrating stochastic step formation corresponding to multiple-soliton states formation with various soliton numbers. No PSC formation observed;
(f,g) Experimentally generated PSC(f) and multiple-soliton state(g), obtained at the powers $P_2$ and $P_4$ correspondingly.
\label{fig_2}}
\end{figure*}

\noindent \textbf{Generation of soliton crystal states} --- 
\label{Generation of SC}
We next focus on the generation of soliton crystals. It has been experimentally demonstrated that soliton crystals can appear as a result of microresonator mode interactions \cite{cole2016solitoncrystal}, which through the avoided modal crossings (AMX) induce a modulation on the intracavity CW background, leading to the ordering of the DKS pulses \cite{wang2017universalbindingDKS}. We would like to point out, however, that to the best of our knowledge all current microresonators inevitably contain AMX-s, which even in a quasi-single-mode case can result from the interaction between fundamental modes \cite{kordts2015higher,grudinin2017high}. The latter means that, in principle, every microresonator system should have the ability to generate a crystal state, because a necessary requirement for crystallization is satisfied. Since the soliton crystals were rarely reported and so far not carefully investigated, it is reasonable to assume that there exists another important ingredient, enabling the formation of soliton crystals, which was not understood yet. We claim, that such a second ingredient is the pump power of the DKS generation procedure, which predetermines the formation of soliton crystal states, and in particular PSC.

In the experiments we observed that the generation of soliton crystal states is typically achieved at relatively low pump powers, while the same standard procedures of soliton excitation (forward tuning \cite{herr2014soliton}) at high pump powers can only lead to the formation of multiple soliton states with a structured spectrum and irregular soliton arrangements. This fact was observed in microresonators having various designs and FSRs, with only a difference in the actual threshold value ($P_{\rm sc}$) distinguishing "low" and "high" pump powers.  To clearly demonstrate the existence of such a threshold, we use four different fixed pump powers ($P_1 - P_4$) and carried out 100 pump frequency sweeps over the cavity resonance in one of our 100-GHz devices. At each pump power the success rate of the generation of PSC states was counted through the statistics of the recorded soliton steps. Figure 2(d) shows the histogram of such success rates for measured pump powers revealing the existence of a clear threshold for deterministic PSC formation at around 0.25W. Strikingly, for both experimental power values below the threshold - $P_1 = 0.15$W and $P_2 = 0.20$W, the system has a long soliton step, which is reproduced in every scan with 100 \% success rate (see Fig.\ref{fig_2}(f)). On this step the system \emph{always} stabilizes in the same PSC shown in Fig.\ref{fig_2}(f), i.e. making the process deterministic. In contrast, reproducing an experiment at higher pump power $P_3 = 0.25$W, i.e. above the threshold, we can observe that soliton steps are stochastically distributed, thus reducing the success rate of the PSC generation. At a high pump power well above the threshold ($P_4 = 0.8$W) the system is purely stochastic in terms of the number of generated soliton pulses (see Fig.\ref{fig_2}(e)), and moreover this number is always well below the maximum soliton number in the PSC state ($X = 15$). 

To reproduce the observed behavior in simulations and understand the underlying physics, we use perturbed Lugiato-Lefever equation (LLE) with the parameters corresponding to our experimental $\rm Si_3N_4$ device (see Supplementary Information for details).  We implement multiple forward tuning scans at various pump powers imitating soliton generation attempts. As in experiments, we also observed the existence of a threshold pump power ($\sim0.25$W, very close to the experimental value), which separates two different generation scenarios. In the first one, below the threshold, \emph{every} simulation ends in the same PSC state, as shown in Fig.\ref{fig_2}(b). The process does not depend on the initial conditions and reveals determinism and extreme robustness of the generation procedure available for the PSC states below the $P_{\rm sc}$. In contrast, the simulation result very quickly becomes stochastic for pump powers above the threshold. Depending on the initial conditions and scan parameters the system forms soliton crystals with defect(s) or -- only in rare cases -- the PSC. At high enough powers no soliton crystal formation is observed. The resulting intracavity waveform is a typical multiple-soliton state, consisting of several sparsely-spaced DKS pulses, as shown in Fig 2(c). Even though the pulses can still maintain long-range ordering (being bound to the modulated background), the characteristic signatures of soliton crystal states (e.g. enhanced lines with extreme conversion efficiency) are degraded.

We found that the observed behavior can be linked to the stability diagram of the LLE (Fig.\ref{fig_2}(a)). Being highly nonlinear, the system has multiple regions with different stability properties, including modulation instability (MI), stable stationary DKS, breathers, spatiotemporal chaos (STC) and transient chaos (TC), which we map out for our system in the second set of simulations (see Supplementary Information). We note, that the LLE can be reduced to the dimensionless form with only two control parameters: normalized detuning $\delta$ and normalized pump amplitude $f$ (see Methods). For this reason the obtained stability diagram and the following discussion can be directly generalized to any Kerr nonlinear microresonator system using dimensionless parameters $(f,\delta)$. We also note that attempts to simulate the stability chart for soliton crystals have been made recently \cite{Karpov2017solitonCrystal,wang2018robust}, but the investigation of its complex structure and the presence of chaotic regimes was incomplete.

By comparing experimental results and simulations of forward scans with the stability diagram, we found that two instability regions - STC and TC play a major role in the formation of soliton crystal states. We discover that the excursion of the system through any of these regions reduces the probability of generating PSC or soliton crystals with a low number of defects.  First, in the region of STC, which has its lower boundary at $P \sim 0.25$W ($f\sim3$), the intracavity waveform experiences fluctuations in the instant number of pulses due to its complex chaotic behavior. Such fluctuations do not guarantee that the number of seed pulses, at the moment when the DKS state is stabilized, will match to the number of potential "sites" introduced by the background modulation. This stochasticity results in fundamental indeterminism in the final DKS state, which can be either a PSC or soliton crystal with defects.
 
The second region -- TC, which reveals even more dramatic impact -- appears above $f\sim 4$ for PSC states, and immediately follows the region of STC for a forward tuning procedure. A prominent feature of this region is that the system there does not have inhomogeneous attractors, and always converges to the trivial, flat solution. This convergence, however, is not immediate and its character and duration depends on the initial conditions. An important consequence of this fact is that any pulsed solution including DKS state or PSC will gradually (pulse-wise) decay to a CW background, while the pump and effective detuning of the state are maintained within the TC region (see more details in Supplementary information). 

The effects of both regions explain our experimental observations and presence of threshold power. Since the generation procedure of the DKS state for any power below $f \sim 3$ ($P_{sc}\sim 0.25$W) avoids the STC - the system deterministically lands in a PSC state. On the other hand when generating a DKS with the pump power above the threshold $f > 3$, the final number of pulses in a DKS state becomes stochastic, and in accordance with experiments the probability of getting PSC decreases. At high pump powers  $f > 4$, when the system experiences the cumulative effect of both STC and TC regions, the formation of the PSC as well as soliton crystals with low defect numbers is prohibited due to the impact of TC, which "clears" the cavity and limits the maximum soliton number of the generated DKS states. 

Our results establish a critical role of the pump power in the generation process of soliton crystals, and more importantly PSC states. They directly provide a simple generation approach ($f<3$) for deterministic access to a PSC state in any microresonator system, which we successfully realized in order to obtain PSC in the microresonators with vastly different FSRs (20-1000 GHz) and fabricated with different fabrication processes (see Fig.\ref{fig_1}(b)). In particular, we were able to generate a perfect soliton crystal state consisting of 87 DKS pulses in a 20-GHz microresonator, which resulted in the interferometric power enhancement of each of its supermodes by almost 40 dB.

\begin{figure*}
\includegraphics[width = \textwidth]{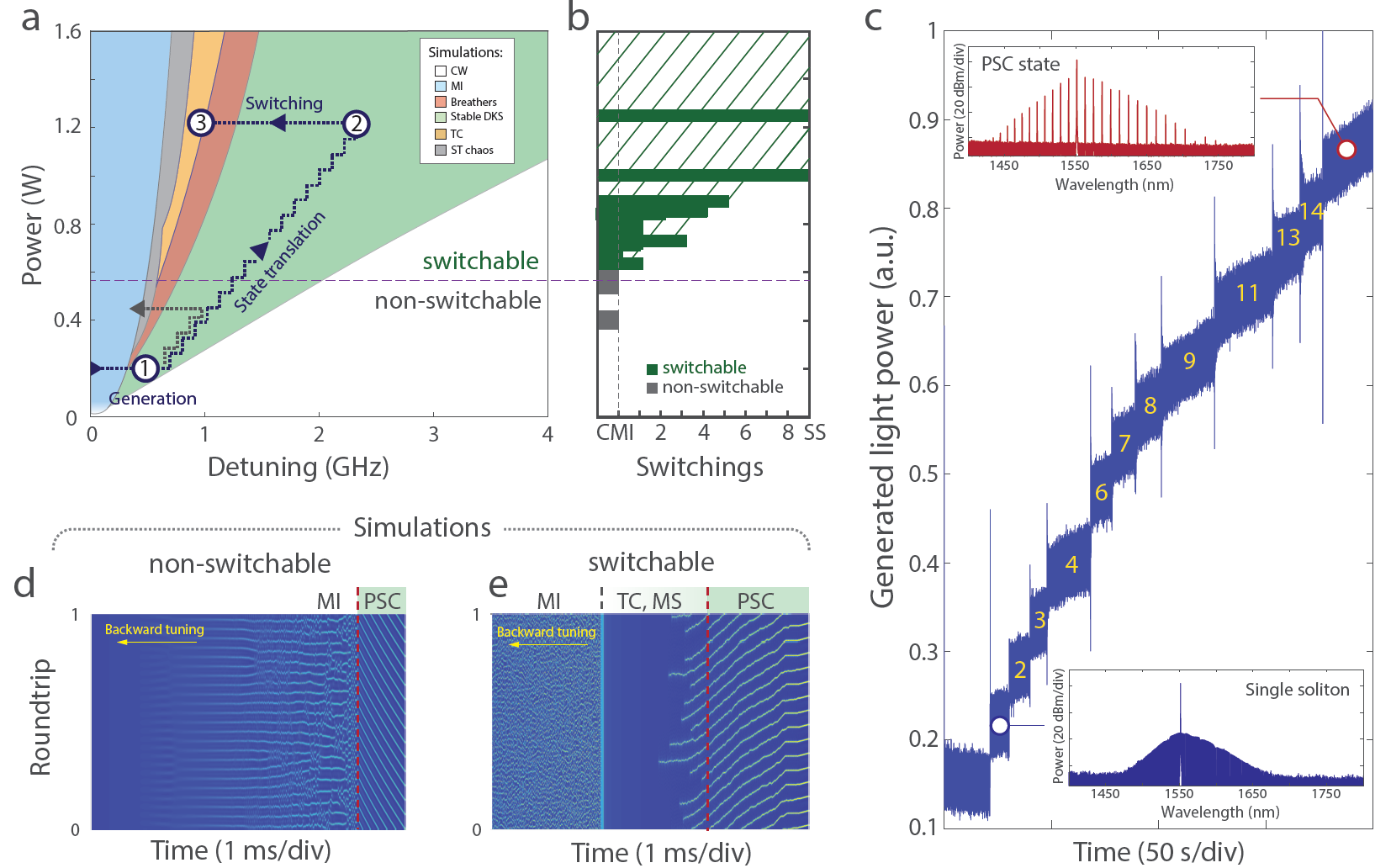}
\protect\caption{\textbf{Controllable translations and switching of PSC states}
(a) Simulated stability chart of the Lugiato-Lefever equation shown in Fig.2(a); Dark blue dashed line traces the complex experimental routes for the controlled PSC state evolution: (1) generation (below TC region), (1 -- 2) power and detuning translation to higher pump power $P'$, (2 -- 3) backward tuning (into TC region) to induce switching. Grey dashed line indicates the route, when the backward tuning is implemented at too low power and system does not enter the TC region, such that no switching is induced.
(b) Experimental results showing the number of switching available to PSC state at different pump powers. The states are unswitchable (grey attempts) below a threshold power of $P_{\rm sw} = 0.6$ W, and are switchable above (green attempts). Horizontal axis shows the number of available switchings, CMI -chaotic modulation instability, SS - switching to single soliton state. We note that the experimentally measured transition between switchable and non-switchable states (marked with purple dashed line) matches well with the bottom of the TC region obtained in simulations shown in (a).
(c) Experimental trace of generated light during the continuous switching of a PSC state to a single-soliton trace. Insets show the optical spectra of the corresponding states.
(d, e) Simulations of the intracavity waveform evolution during backward tuning of the PSC at non-switchable pump power (g) and at switchable pump power (h). In the first case the system directly seeds MI, while in the second case the system first decays to a flat solution (CW) via multiple-soliton states (MS).
\label{fig_3}}
\end{figure*}

\noindent \textbf{Switching of soliton crystal states} ---
\label{Switching of SC + method}
After deriving the conditions for the faultless and deterministic generation of PSC states we focus on their dynamical properties. We start by introducing a method to operate with soliton states in the two-dimensional parameter space of the system (pump power, effective detuning) in order to access different dynamical regimes of soliton crystals. Using the recently developed soliton probing scheme \cite{karpov2016universal} we are able not only to determine the current state of the system, but precisely route it in the parameter space. Employing this method, we experimentally verified that PSC states are robust with respect to the various power- and detuning-translations, and are maintained by the system until its parameters stay within the soliton existence area. This gives us a powerful opportunity to explore different dynamical regimes of soliton crystals by implementing arbitrary complex routes of PSC state transfer in power, detuning, or both.

In order to demonstrate the capabilities of the method, we use it to investigate the switching process of soliton crystal states. Indeed, the switching of soliton crystal states has never been reported, though a recent work has shown that DKS states can experience switching events, which reduce the number of intracavity solitons one-by-one \cite{karpov2016universal}. Here we have first experimentally realized the switching behaviour for soliton crystal states, and second - more fundamentally - we have discovered an origin of the DKS switching ability. 

\begin{figure*}
\includegraphics[width = \textwidth]{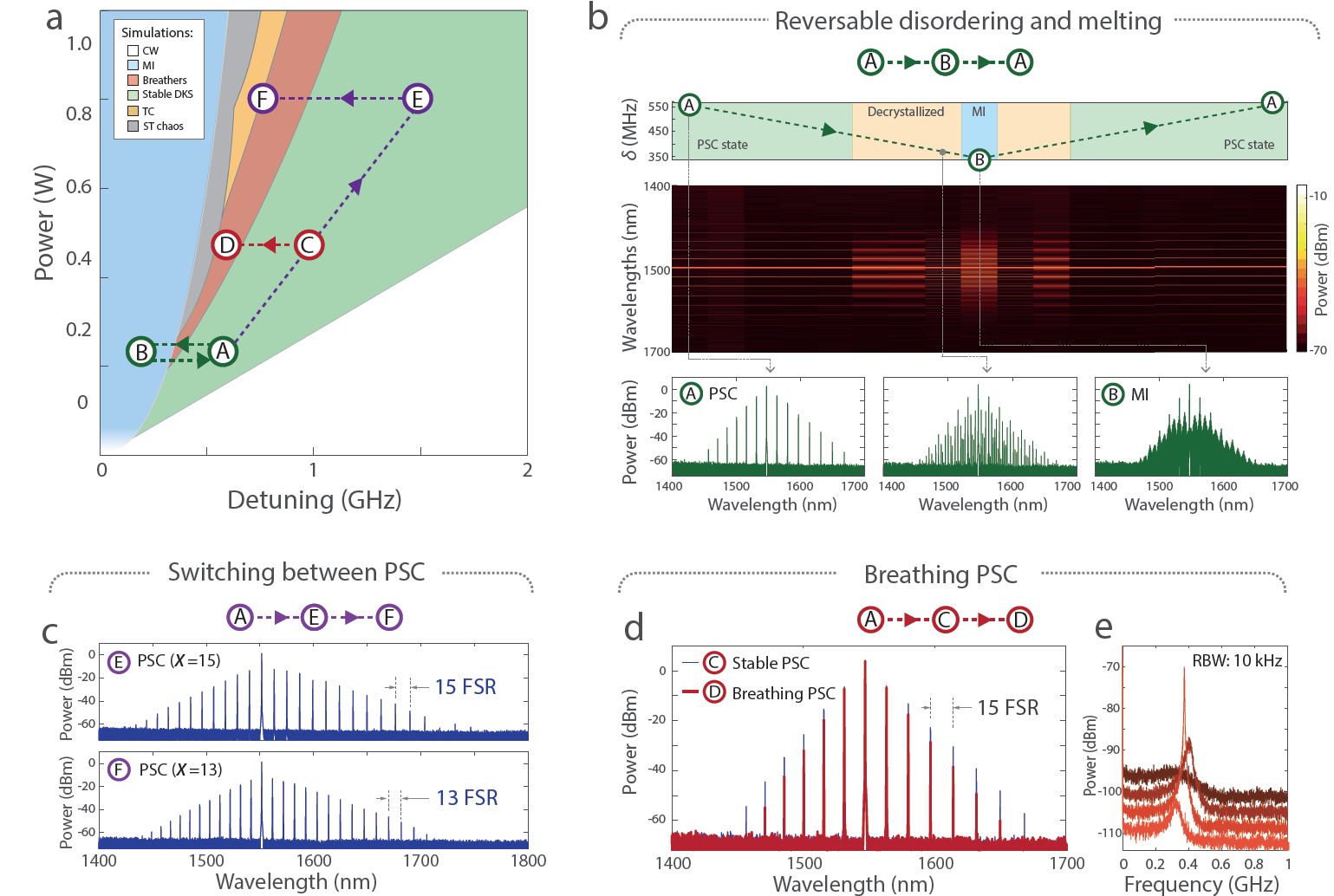}
\protect\caption{\textbf{Diverse dynamics of perfect soliton crystals states}
(a) Simulated stability chart of the Lugiato-Lefever equation shown in Fig.2(a). Three experimental evolution routes of the initial PSC state (A) are marked with different colors. 
(b) Experimental observation of PSC melting and recrystallization. The route is shown as A -- B -- A in (a), and passes below STC and TC regions. The detuning was reduced linearly until the system has reached modulation instability (MI), and then increased to the starting value, as shown at the top plot. Middle map shows the evolution of the optical spectrum of a PSC state during excursion to the MI regime (crystal melting) and then back to the PSC (recrystallization after melting). The bottom plot shows three spectra at different stages, from left to right: initial PSC state, disordered soliton crystal state, and modulation instability at the corresponding stages of the tuning route.
(c) Switching between perfect soliton crystal states. The route is shown as A -- E -- F in (a). Optical spectra of the initial perfect soliton crystal state with $X=15$ and final perfect soliton crystal state with $X = 13$, obtained "on the fly" by power translation and switching from the initial.
(d) Optical spectra of the stable (blue) and breathing (red) PSC states. Both states maintain strict regularity of the DKS pulses.
(e) Evolution of the total intensity noise spectrum as the PSC is tuned into the breathing region. The appearance and reduction of the sharp tone, corresponding to the breathing frequency can be observed. The spectra with larger detuning have darker color, the traces are shifted vertically by 5 dB each for better visualization of the evolution.
\label{fig_4}}
\end{figure*}  

We constructed complex routes consisting of three consecutive stages (see Fig.\ref{fig_3}(a)): (1) PSC generation with forward tuning at the power below $P_{\rm sc}$, (2) PSC state translation to a new power $P'$, (3) backward tuning of the state in an attempt to induce switching. Implementing such routes for various $P'$ values in the same 100-GHz $\rm Si_3N_4$ microresonator as in the previous section, we observed that the switching of PSC states suddenly becomes available above a pump power $P_{\rm sw} \sim 0.6$ W $(f_{\rm sw} \sim 4)$. Below $P_{\rm sw}$ the system does not show any switching behaviour, and the PSC state directly seeds MI (a similar effect was observed for individual DKS in fiber cavities \cite{anderson2016chaos}). Above the $P_{\rm sw}$ as the pump power is increased, the number of available switchings increases until the system is able to reach a CW solution before the transition to the MI state (above $\sim 1$ W $(f=5)$). In most cases at intermediate power, the direct MI seeding can still be observed, but happens after several switchings. Figure 3(c) shows an example of such behaviour, where the system was switched from a PSC state with 15 DKS pulses to a single soliton state, and then seeded MI.

In order to verify this behaviour in simulations, we used the same LLE as in the previous section, and implemented backward tuning procedures of PSC states at various powers. Similarly to the experiments, we observed that the PSC state starts to experience switching behaviour above the pump power of 0.55W, which is very close to the experimentally obtained $P_{\rm sw}$. Comparing this threshold to the stability diagram, one can see that, strikingly, it essentially coincides with the lower boundary of the TC region in the stability diagram obtained for the PSC states. Below the TC region, all simulated PSC states are non-switchable, and directly seed MI in the backward tuning, as shown in Fig.\ref{fig_3}(d). In contrast, when the system's route passes the region of TC, one, several or all DKS pulses can decay (see Fig.\ref{fig_3}(e)). These observations let us establish a fundamental link between the soliton switching ability and the TC regime of driven nonlinear microresonators. Moreover, such a link reveals the underlying reason of soliton switching, which is in principle valid not only for a PSC state but for any DKS as well: when the stable DKS state is translated into the TC region, the system starts to spontaneously lose pulses from the cavity. However once the first pulse is decayed, the thermal effects of cavity cooling shift the system back to the stable soliton existence range \cite{karpov2016universal}.
At this point, we would like to bring the readers attention to two recent studies \cite{wang2018robust,lu2018deterministic}, which experimentally demonstrated the generation of soliton crystal states and DKS switching in thermally-controlled hydex microresonators, and whose observations can be well explained using the understanding of PSC generation and switching developed in the present work (see Supplementary information).

\noindent \textbf{Dynamics of soliton crystal states} ---
\label{Dynamics, breathing, recrystallization}
In the final section, we would like to uncover a rich panel of peculiar dynamical phenomena, which can be found in PSC states and studied using our methods developed above. We highlight three of them, which were observed, using different routes of the PSC state translations (shown in Fig.\ref{fig_4}(a): reversible melting and recrystallization of the PSC state, switching between PSC states and the formation of PSC breathers.

First, we demonstrate that a PSC state can be consistently restored after its excursion to the MI region. This corresponds to a complete destruction of the regular soliton arrangement - ``soliton crystal melting", and its reassembling, when the system is brought back to the region of stable DKS -  ``soliton recrystallization" (see A--B--A route, marked with green in Fig.\ref{fig_4}(a)). 
The initial PSC state (A) was tuned backward until it seeds MI (A -- B), and then - forward (B -- A) to the initial position. We trace the system evolution during this procedure by measuring the optical spectrum. For the major part of the stable DKS region, the system maintains a typical PSC spectrum. It starts to develop additional lines apart from the supermodes as the system approaches the MI region indicating the appearance of variations in the relative positions of DKS pulses, while maintaining the overall long-range ordering. The effect is similar to the introduction of disorder in the crystal lattice of solids. Once the system reaches MI - the spectrum changes to the typical spectrum of a noisy comb and the system acquires strong intensity noise. In this state stable DKS pulses cannot exist in the system, and the intracavity waveform is chaotic, meaning that there is no order in the time domain. Reverting the tuning direction and bringing the system back to the initial state (B -- A), we observe that the system can be restored back to the initial PSC state, where the DKS pulses are again crystallized in the form of an equidistant lattice. Reproducing the described tuning in the simulation, we can observe the described evolution of the intracavity waveform (see Supplementary information). We note that the conversion of certain SC states to MI and back has been also recently accessed in a non-deterministic fashion, with non-deterministic restoration of the initial soliton crystal \cite{wang2018robust}, and was linked to the chaotic behaviour in the MI state.

Second, we demonstrate that in certain cases, the system can be switched from one PSC state to another PSC state with a distinct number of intracavity pulses (see A - E - F route in Fig.\ref{fig_4}(a)). In our experiments it was enabled by a proper choice of the pump power (above $P_{sw}$) at which the system was switched, and led to the switching from a PSC with $X=15$ to the one with $X=13$ (see Fig.\ref{fig_4}(c)). We attribute this dynamic to the change in the modulation of the CW background, which could be caused by the cavity cooling after switching. Since the positions and strength of the modal crossings are sensitive to the temperature of the system \cite{xue2015darksol,xue2015AMXcontrol}, the removal of the DKS pulses can change it and induce a new binding potential with a different number of sites.

Third, we experimentally demonstrate the formation of perfect soliton crystal breathers, which correspond to simultaneous oscillations in the amplitude and duration of all DKS pulses forming the PSC state. For this we brought the PSC state to the breathing region (see the route A - C - D, marked red in Fig.\ref{fig_4}(a)), where the characteristic indicators of the breathing DKS states have been observed, including the triangle-shaped optical spectrum, and the appearance of the narrow breathing tone, whose frequency was close to the estimated effective detuning and was decreasing as the detuning decreased (see Fig.\ref{fig_4}(d)) \cite{lucas2017breathing}.

\label{Discussion}
\noindent \textbf{Discussion} ---
We have demonstrated platform-independent on-demand generation of \emph{perfect soliton crystals}, which essentially represent defect-free soliton lattices.
In comparison to other soliton crystal states (with defects), they show maximized conversion efficiency into the supermodes spaced by $X\cdot$FSR, whose power moreover scales as $X^2$. First, both in experiments and LLE-based simulations, we discovered that the generation of PSC states is highly sensitive to the pump power, which results from the complex structure of the systems stability diagram including the regions of STC (perturbing the PSC state and inducing soliton interactions) and TC (stochastically clearing the intracavity DKS pulses). Based on this understanding we derived and verified a deterministic, device-independent procedure for the generation of PSC states, which simply requires the use of normalized pump power $f<3$. Second, we have demonstrated that the region of TC has a significant impact not only on the process of the soliton generation, but constitutes a fundamental feature of the driven-dissipative nonlinear cavity enabling switching of the DKS states. By translating the PSC states above the normalized power of $f\sim4$ we managed to make them switchable and even demonstrated a consecutive transition to a single state. Taking into account the two important thresholds for the generation of PSC states ($f_{\rm sc}\sim3$) and their switching ($f_{\rm sw}\sim4$), one can in principle construct deterministic routes to access any soliton state available in the system.
Finally, we demonstrated the rich dynamics of the PSC states, including their ability to sustain translation to the MI state and form the same PSC state afterwards, switching between soliton crystals and formation of the PSC breathing states.

All demonstrated PSC states are easily accessible and enable lossless energy redistribution in particular optical modes, which can be of high use for the development of a chip-scale optical frequency combs sources with ultra-high-repetition rates beyond several THz. We also believe that such states can provide a convenient microwave-to-THz link, enabling the stabilization of THz signals with standard RF equipment.

We also would like to emphasize that the above-listed results showing the impact of chaotic regions on the formation and switching processes of solitons states as well as the translation methods we developed to study the complex dynamics of the system can be extended for any DKS states, including less temporally organized multiple-soliton states or single DKS. 
Our observations are establishing critical links between the fundamental stability chart of the CW-driven nonlinear microcavity and the formation and the switching dynamics of the DKS states.  We strongly believe that such a clear understanding of the effect of TC on these and other processes will help to uncover and explain many other phenomena of the DKS states in optical microcavities including connection to recent theories and explanation of previous observations \cite{karpov2016universal,joshi2016thermal,cole2016solitoncrystal,wang2018robust,lu2018deterministic,sun2018stability}.

\section*{Methods}
\label{Methods} 
\noindent\textbf{Optical resonators}. $\mathcal{\mathcal{\mathrm{Si_{3}N_{4}}}}$
integrated microring resonators with the free spectral range (FSR) of $\sim100$ $\mathrm{GHz}$ and Q-factors $\sim10^6$ (linewidth $\frac{\kappa}{2\pi}=150-200~{\rm~MHz}$)  were fabricated using the
Photonic Damascene process \cite{Pfeiffer2015Damascene}. In order to achieve single mode operation and suppress the effect of avoided mode crossings, a ``filtering section'' was added to the microresonator \cite{Herr2014avoidmodecross,kordts2015higher}. Similarly, the photonic Damascene process was employed for the fabrication of the 200-GHz, and 1-THz devices used for the verification of the results presented in the paper, and in particular for demonstrating the platform-independent generation of PSC. Also, 20-GHz microresonators, where we demonstrated the largest to date soliton crystal consisting of 87 regularly spaced solitons, have been fabricated with a typical subtractive process.

\noindent\textbf{Dimensionless form of the LLE}.
The dimensionless form of the LLE can be written for the normalized intracavity waveform $\Psi(\tau,\theta)$:
		
\begin{equation}
i\frac{\partial\Psi}{\partial\tau}+\frac{1}{2}\frac{\partial^{2}\Psi}{\partial\theta^{2}}+|\Psi|^{2}\Psi=(-i+\zeta_{0})\Psi+if.\label{eq:nls}
\end{equation}
		
Where $\theta$ is the dimensionless longitudinal coordinate,  and $\tau$ is the normalized time. The system's dynamics are determined by two main parameters: the normalized pump power $f^2$ and effective detuning $\zeta_0$, defined as \cite{herr2014soliton}:
\begin{equation}
\begin{aligned}
		    f^2 &= \dfrac{8 g \eta P_{\rm in}}{\kappa^2 \hbar \omega_0} \ , & \zeta_0 = \dfrac{2\delta\omega}{\kappa} \ ,
\end{aligned}
\end{equation}
\noindent
where $\kappa$ denotes the total resonator linewidth ($Q=\omega_{0}/\kappa$, loaded quality factor), $\eta=\kappa_{\mathrm{ex}}/\kappa$ the coupling coefficient, $P_{\rm in}$ is the pump power, $\omega_{0}$ the pumped resonance frequency and $\delta\omega = 2\pi\delta = \omega_{0}-\omega_{p}$ is the detuning of the pump laser from this resonance, counted positive for a red-detuned laser. The nonlinearity is described via $g=\hbar\omega_{0}^{2}cn_{2}/n_{0}^{2}V_{\mathrm{eff}}$ giving the Kerr frequency shift per photon, with the effective group refractive index $n_{0}$, nonlinear refractive index $n_{2}$, and the effective optical mode volume $V_{\mathrm{eff}}$.

\noindent \textbf{Acknowledgements}
\begin{acknowledgments} We gratefully acknowledge fruitful discussions with Erwan Lucas and Miles Anderson.
This publication was supported by the Air Force Office of Scientific Research, Air Force Material Command, USAF under Award No. FA9550-15-1-0099; 
by funding from the European Union's Horizon 2020 Marie Sklodowska-Curie IF grant agreement No. 753749 (SOLISYNTH);
and by the Swiss National Science Foundation under grant agreement No. 161573 (precoR);
This publication was supported by contract D18AC00032
(DRINQS) from the Defense Advanced Research Projects Agency, Defense Sciences
Office;
M.K. acknowledge the support from  the European Space Technology Centre with ESA Contract No. 4000116145/16/NL/MH/GM.
${\rm Si_3N_4}$ samples were fabricated and grown in the Center of MicroNanoTechnology (CMi) at EPFL.
\end{acknowledgments}

\noindent \textbf{Author contributions} \\M.K. developed the idea, designed and performed experiments and simulations, and processed the data. M.H.P.P. fabricated samples with the assistance of J.L.. M.K. wrote a manuscript with input from T.J.K., H.G., W.W.. T.J.K. supervised the project.


%

\end{document}